\documentclass[superscriptaddress,amsmath,amssymb,10pt,pra,floatfix,twocolumn,longbibliography]{revtex4-2}

\usepackage{xcolor}
\usepackage{graphicx}
\usepackage{float}
\usepackage{bm}
\usepackage{hyperref}

\begin{document}

\title{
Josephson spectroscopy in a circular atomic tunnel junction with acceleration-induced symmetry breaking
}

\author{Yurii Borysenko}
\affiliation{Department of Physics, Taras Shevchenko National University of Kyiv, 64/13, Volodymyrska Street, Kyiv 01601, Ukraine}
\affiliation{Technische Universität Graz, Rechbauerstraße 12, 8010 Graz, Austria}

\author{Yuriy Bidasyuk}
\affiliation{Bogoliubov Institute for Theoretical Physics, 14-b Metrolohichna Street, Kyiv 03143, Ukraine}

\author{Olena Prykhodko}
\affiliation{Department of Physics, Taras Shevchenko National University of Kyiv, 64/13, Volodymyrska Street, Kyiv 01601, Ukraine}

\author{Gerhard Birkl}
\affiliation{Technische Universität Darmstadt, Institut für Angewandte Physik, Schlossgartenstraße 7, 64289 Darmstadt, Germany}
\affiliation{Helmholtz Forschungsakademie Hessen f\"ur FAIR (HFHF), GSI Helmholtzzentrum für Schwerionenforschung, 64291 Darmstadt}

\author{Dominik Pfeiffer}
\affiliation{Technische Universität Darmstadt, Institut für Angewandte Physik, Schlossgartenstraße 7, 64289 Darmstadt, Germany}

\author{Ludwig Lind}
\affiliation{Technische Universität Darmstadt, Institut für Angewandte Physik, Schlossgartenstraße 7, 64289 Darmstadt, Germany}

\author{Mark Edwards}
\affiliation{Department of Biochemistry, Chemistry, and Physics, Georgia Southern University,
Statesboro, Georgia 30460-8031, USA}

\author{Alexander Yakimenko}
\affiliation{Department of Physics, Taras Shevchenko National University of Kyiv, 64/13, Volodymyrska Street, Kyiv 01601, Ukraine}
\affiliation{Dipartimento di Fisica e Astronomia Galileo Galilei,
Universit\'a di Padova,
Via Marzolo 8, 35131 Padova, Italy}

\begin{abstract}

\textcolor{black}{We study Josephson dynamics in a long atomic Bose-Josephson junction formed by two tunnel-coupled coplanar Bose-Einstein-condensate rings. An in-plane linear acceleration breaks the axial symmetry of the trap and transforms a single Josephson plasma oscillation into a multimode population-imbalance response. Gross-Pitaevskii simulations and Bogoliubov-de Gennes analysis show that the additional spectral components arise from collective modes that acquire finite overlap with the population-imbalance operator under symmetry breaking, with their activation governed by reflection symmetry about the acceleration direction.
We also propose a mode-resolved Josephson-spectroscopy protocol based on a weak localized periodic perturbation. Frequency scans reveal resonant amplitude peaks and phase shifts at the eigenfrequencies of active Bogoliubov modes, while angular scans of the drive position provide access to the angular structure of the corresponding mode density perturbations. A dissipative time-dependent Bogoliubov theory yields analytical response functions in quantitative agreement with full Gross-Pitaevskii simulations in the linear regime. Our results demonstrate that accelerated dual-ring condensates provide a controllable platform for symmetry-selected Josephson dynamics and spectroscopic probing of collective modes.
}
\end{abstract}

\maketitle

\section{Introduction}
The Josephson effect is among the most fascinating phenomena of quantum physics accessible for detailed investigation in atomic Bose-Einstein condensates (BEC) \cite{Smerzi1997, Albiez2005, Levy2007, Gati2007, eckel2014hysteresis, PhysRevLett.111.205301, PhysRevLett.113.045305, Valtolina2015}. It manifests itself in the coherent tunneling of superfluid matter through a potential barrier separating two weakly coupled condensates, leading to characteristic oscillations of the population imbalance and relative phase between them. 
The Josephson effect has been extensively studied in various configurations of BECs, including double-well potentials with a point tunnel barrier \cite{Smerzi1997, Albiez2005}, ring-shaped traps \cite{eckel2014hysteresis, PhysRevLett.111.205301}, parallel tunnel-coupled condensate waveguides \cite{PhysRevLett.106.025302, momme2019collective}, and more complex geometries \cite{oliinyk2019tunneling, PhysRevA.111.043308}.

The frequency of small-amplitude Josephson oscillations is determined by the energy difference between symmetric and antisymmetric states of the coupled condensates, which in turn depends on the tunneling rate through the barrier and the interaction strength within each condensate. In simple geometries, the Josephson frequency is well understood and can be accurately predicted by effective theories such as the two-mode model \cite{Smerzi1997, raghavan1999coherent}. However, in more complex geometries or in the presence of external perturbations, the tunneling dynamics can become significantly more intricate, leading to deviations from the two-mode picture and requiring more sophisticated theoretical approaches
\cite{ananikian2006gross, momme2019collective}.

In extended and multiply connected condensate geometries, Josephson tunneling is naturally accompanied by additional collective degrees of freedom, so that the population-imbalance dynamics can no longer be reduced to a single plasma mode. Ring-shaped systems \cite{eckel2014hysteresis, PhysRevLett.111.205301,Turpin2015} are especially well suited for this regime, since they combine weak-link-mediated transport with a closed superfluid geometry in which symmetry and collective excitations play a central role \cite{pitaevskii2016bose,Amico2021,PhysRevLett.106.130401,PhysRevLett.111.205301,Wang2015}. 
Previous studies have shown that, beyond the simplest two-mode picture, Josephson oscillations may hybridize with phononic and Bogoliubov excitations, leading to resonant and multimode dynamics in spatially extended junctions \cite{bouchoule2005modulational,PhysRevA.94.033603,momme2019collective,PhysRevA.95.023627,PhysRevA.98.043624,Gallemi2015,PhysRevA.102.043316,Paraoanu_2001,doi:10.1143/JPSJ.74.3179}. Related transport phenomena in coupled ring condensates, including persistent-current tunneling and coherent current transfer, have also been explored in recent years \cite{oliinyk2019tunneling,PerezObiol2022,Bland2022}. This motivates the study of a dual-ring Bose Josephson junction under controlled symmetry breaking, where the interplay between tunneling and collective modes can be investigated in a particularly clear form. \textcolor{black}{In this setting, the external perturbation provides a tunable way to control the overlap between collective modes and the observable population-imbalance.}

\textcolor{black}{In this work we investigate Josephson population-imbalance dynamics in a long atomic Bose-Josephson junction formed by two concentric superfluid rings separated by a potential barrier. The same coplanar dual-ring geometry was recently used to study acceleration-driven Josephson-vortex dynamics, where linear acceleration produces asymmetric tunneling flows, vortex displacement, and signatures relevant for inertial sensing~\cite{PhysRevA.111.043308}. Here we address a different aspect of the same platform: the coupling between Josephson population-imbalance oscillations and collective Bogoliubov modes in the absence of imposed persistent-current mismatch. Using numerical simulations of the Gross-Pitaevskii equation (GPE) and Bogoliubov-de Gennes (BdG) analysis, we show that an in-plane linear acceleration transforms the single Josephson oscillation of the axially symmetric system into a multimode response. The additional spectral components are not interpreted as the simple splitting of one Josephson eigenmode, but as the activation of specific Bogoliubov modes that acquire finite overlap with the population-imbalance operator when rotational symmetry is reduced. We further propose a mode-resolved Josephson-spectroscopy protocol based on a weak localized periodic perturbation. When the drive frequency is tuned near an active Bogoliubov eigenfrequency, the population-imbalance response exhibits a resonant enhancement and phase shift, while angular scans of the drive position provide access to the spatial structure of the corresponding mode.}
\textcolor{black}{Recent cavity-optomechanical proposals have also shown that ring-condensate states, including persistent currents, solitons, and Josephson dynamics, can be monitored in situ and with minimal destruction, providing complementary readout strategies for atomtronic circuits~\cite{Kumar2021PRL,Pradhan2024PRR,Pradhan2025PRR}.}  

The paper is organized as follows. In Section II, we introduce the theoretical model and the numerical methods used in our study. In Section III, we analyze the effect of external acceleration on the Bogoliubov spectrum and its relation to the Josephson oscillation frequencies. 
In Section IV, we investigate the resonant excitation of Bogoliubov modes using an external periodic perturbation. We demonstrate how Josephson oscillations can be utilized as a sensitive probe for detecting and characterizing these modes. 
Finally, in Section V, we summarize our findings and discuss their implications for future research on Josephson dynamics in double-ring BECs.

\section{Model}

We consider a  BEC of $^{87}$Rb atoms confined in a coplanar coaxial double-ring trap with a repulsive potential barrier separating the inner and outer rings, as shown in Fig.~\ref{fig:1_setup}.
In cylindrical coordinates $\mathbf{r}=(r,\varphi,z)$ the double-ring potential can be parametrized as follows:
\begin{figure}[tbp]
    \includegraphics[width=\linewidth]{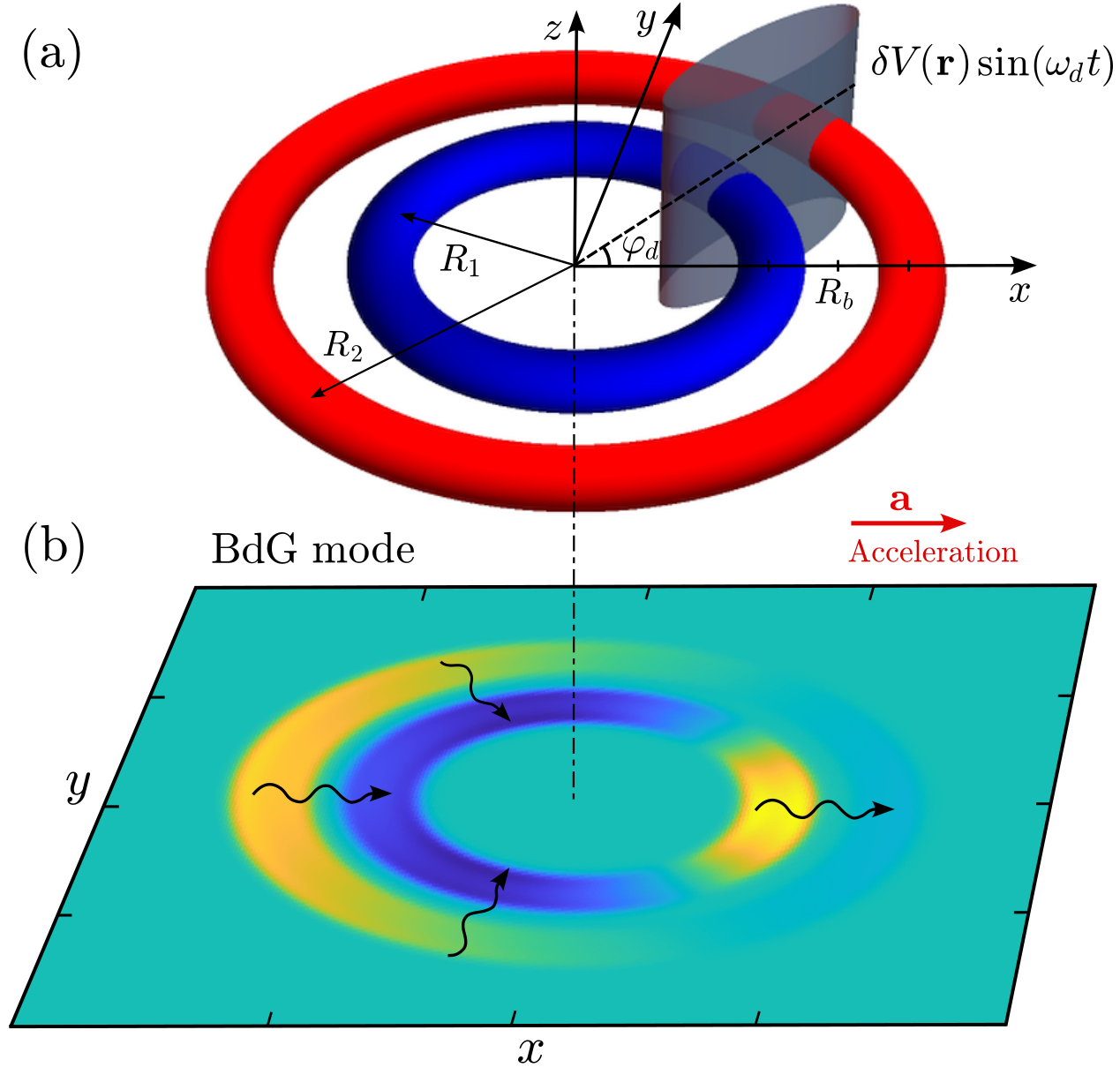}
    \caption{Schematic of the dual-ring Bose Josephson junction. (a) Two coaxial toroidal condensates separated by a barrier that supports coherent tunneling between the rings, together with an additional time-dependent potential used to perturb the system. (b) The density modulation of a Bogoliubov mode that becomes active in the population-imbalance dynamics. Arrows indicate the direction of the tunneling flow across the barrier.}
    \label{fig:1_setup}
\end{figure}
\begin{equation}
    V_\textrm{trap}(\mathbf{r}) = \frac{1}{2}M\omega_z^2 z^2 + V_{\rm dr}(r),
\end{equation}
where $\omega_z=2\pi\times 245$ Hz is the trapping frequency in a light sheet confining the atoms along the $z$ direction.
The radial double-ring potential is given by
\begin{equation}\label{eq:Vdr}
V_{\rm dr}(r) = V_{\rm ring}(r) +V_{\rm b}(r).
\end{equation}
The term $V_{\rm ring}(r)$ defines a ring-shaped potential with a flat region between the two characteristic radii:
\begin{equation}
V_{\rm ring}(r)=V_1(r)\Theta(R_1-r)+V_2(r)\Theta(r-R_2),
\label{eq:Vring}
\end{equation}
where the Heaviside step functions, $\Theta(r)$, ensure that $V_{\rm ring}(r)=0$ for $R_1<r<R_2$, while $V_{\rm ring}(r)=V_1(r)$ for $0\leq r\leq R_1$ and $V_{\rm ring}(r)=V_2(r)$ for $r\geq R_2$. The potentials defining the inner and outer wells, $V_j(r)$ for $j=1,2$, take the form
\begin{equation}
V_{j}(r)=\frac{1}{2}M\omega^{2}_{r}(r-R_{j})^2,
\end{equation}
with identical trapping frequencies $\omega_r = 2\pi\times 110$ Hz and distinct ring radii $R_{1}=14\,\mu$m, $R_{2}=24\,\mu$m.

The term $V_{\rm b}(r)$ represents the barrier separating the rings,
\begin{equation}
V_{\rm b}(r)=U_{\rm b}\exp\left[-\frac{(r-R_{\rm b})^2}{2l_{\rm b}^2}\right],
\end{equation}
with the radius $R_{\rm b}=(R_1+R_2)/2 = 19\,\mu$m and width $l_{\rm b}=1\,\mu$m. 
In the present study we fix the barrier height to $U_{\rm b}=1.7\mu$, where $\mu$ is the chemical potential of the condensate.
The above parametrization of the double-ring potential was previously adopted in \cite{PhysRevA.111.043308} and allows us to independently control the ring radii, as well as the amplitude and width of the barrier separating them.

External linear acceleration is introduced as an inertial force in a reference frame that moves with the system. It is defined by the potential
\begin{equation}
    V_{\rm a} = M(\mathbf{a} \cdot \mathbf{r}),
    \label{eq:Va} 
\end{equation}
with $M$ being an atom mass.
This is the only term in the Hamiltonian that explicitly breaks the axial symmetry of the system. In the present study, we consider acceleration along the $x$ direction, \textcolor{black}{$\mathbf{a} = (a_x,a_y,a_z) = (a,0,0)$}, with the magnitude within the range $0 \leq a \leq 20$ mm/s$^2$.

In the present study we are interested in the low-energy dynamics of the condensate, well below the characteristic excitation energy along the tightly confined $z$ direction. Therefore we can consider the dynamics in the $z$ dimension as frozen and use an effective two-dimensional (2D) description of the system in the $(r, \varphi)$ plane. The dynamical properties of a BEC within the mean-field theory at the zero-temperature limit is governed by the Gross-Pitaevskii equation: 
\begin{equation}
i\hbar\frac{\partial\psi(\mathbf{r},t)}{\partial t}= \left[\Hat{H}_0 + g |\psi(\mathbf{r},t)|^{2}\right]\psi(\mathbf{r},t).  
\label{eq:GPE}
\end{equation}
where  
\[
\Hat{H}_0 = -\frac{\hbar^2}{2M}\nabla_{r,\varphi}^2 + V_\textrm{trap}(r) + V_{\rm a}(\mathbf{r})
\]
is the single-particle Hamiltonian, 
\[
g = \frac{4\pi a_{s}\hbar^{2}}{M} \sqrt{\frac{M\omega_z}{2\pi\hbar}}
\]
is the two-dimensional coupling constant, $M=1.44\times 10^{-25}$ kg and  $a_{s} = 5.3\times 10^{-9}$ m are the mass, and the $s$-wave scattering length of $^{87}$Rb atoms. The wave function is normalized to the total number of atoms in the system: 
\begin{equation}
\int \left|\psi(\mathbf{r},t)\right|^2 d \mathbf{r} = N = 5\times10^4.
\end{equation}

The total number of atoms is distributed between inner and outer rings $N=N_1+N_2$, where:
\begin{equation}
N_j (t)=\iint_{S_j}|\psi(\mathbf{r},t)|^2d\mathbf{r},
    \label{eq:Nequlibrium}
\end{equation}
with integration boundaries for the inner ring, $S_1$: ${0\le r<R_b}$ and for the outer ring, $S_2$: ${r \ge R_b}$. 

We define the population imbalance between these parts as the deviation in the number of particles from their equilibrium values, $N_j^{(0)}$, as follows:
\begin{equation}
 Z(t) = \left[N_2(t)-N_2^{(0)}\right] - \left[N_1(t)-N_1^{(0)}\right].  
    \label{eq:Delta_N}
\end{equation}
Alternatively, the population imbalance can be expressed in an operator form as
\begin{equation}
    Z(t) = \langle \psi(\mathbf{r},t) | \hat{Z} | \psi(\mathbf{r},t) \rangle - \langle \psi_0(\mathbf{r}) | \hat{Z} | \psi_0(\mathbf{r}) \rangle,
    \label{eq:population_imbalance_operator}
\end{equation}
where $\hat{Z} = \Theta(r-R_b) - \Theta(R_b - r)$ is the population imbalance operator, and $\psi_0(\mathbf{r})$ is the stationary solution of the GPE.

The low-energy collective excitations in the condensate can be analyzed by applying a small perturbation to $\psi_0(\mathbf{r})$

\begin{equation}\label{eq:psi_decomp}
   \psi(\mathbf{r},t) = e^{-i \mu t/\hbar} \left[\psi_0(\mathbf{r}) + \delta\psi(\mathbf{r},t)\right],
\end{equation}
where
\begin{equation}\label{eq:delta_psi}
   \delta\psi(\mathbf{r},t) = u(\mathbf{r})e^{-i\omega t}+v^*(\mathbf{r})e^{i\omega t}.
\end{equation}
After substituting Eq.~\eqref{eq:psi_decomp} into the GPE and linearizing with respect to the small perturbation $\delta\psi$, we obtain the \textcolor{black}{Bogoliubov-de Gennes}  equations for the mode amplitudes $u(\mathbf{r})$ and $v(\mathbf{r})$ and their corresponding frequencies \textcolor{black}{$\omega$:
\begin{equation}
\begin{cases}
    \begin{aligned}
        \hbar\omega \, u(\mathbf{r}) &= \hat{\mathcal{L}} u(\mathbf{r}) + g\,\psi_0^2(\mathbf{r})\,v(\mathbf{r}), \\
        -\hbar\omega \, v(\mathbf{r}) &= \hat{\mathcal{L}} v(\mathbf{r}) + g\,\psi_0^2(\mathbf{r})\,u(\mathbf{r}).
    \end{aligned}
    \label{eq:BdG_equations}
\end{cases}
\end{equation}
where 
\begin{equation}\label{eq:L_operator}
\hat{\mathcal{L}} = \Hat{H}_0 - \mu + 2g\left|\psi_0\right|^2.
\end{equation}}

The set of solutions of the BdG equations forms a complete basis of collective excitations, which can be used to analyze the low-energy dynamics of the system and its response to external perturbations.

\section{Josephson oscillations and Bogoliubov modes under external acceleration}

In this section, we investigate how external linear acceleration modifies the collective dynamics of the double-ring Bose-Josephson junction through its coupling to Bogoliubov excitations. We demonstrate that acceleration-induced symmetry breaking produces a qualitative modification of the Josephson spectrum, manifesting as multi-frequency Josephson oscillations that can be understood through an activation mechanism of low-lying collective modes.

\subsection{Josephson dynamics under external acceleration}

\begin{figure*}
    \centering
    \includegraphics[width=\textwidth]{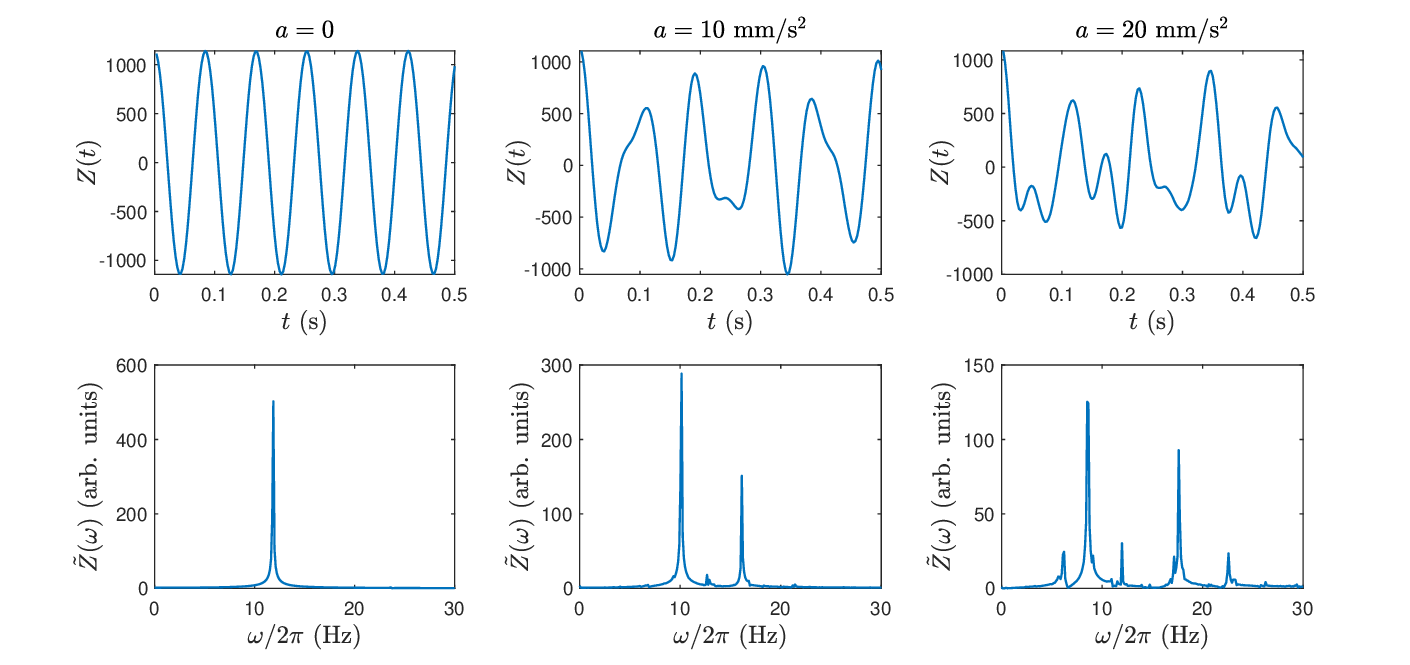}
    \caption{Time evolution of the population imbalance $Z(t)$ (top row) and the corresponding Fourier spectra (bottom row) for three values of the acceleration: $a=0$ (left column), $a=10\,\mathrm{mm/s^2}$ (middle column), and $a=20\,\mathrm{mm/s^2}$ (right column). For a nonzero acceleration, $Z(t)$ exhibits a pronounced beating pattern, indicating the presence of several active oscillation frequencies. In the Fourier spectra, the single dominant Josephson peak observed at $a=0$ evolves into a multiplet structure whose components shift with increasing acceleration.}
    \label{fig:2_N_and_FFT}
\end{figure*}

To initiate Josephson oscillations in our system, we apply a small quench potential that creates a population imbalance between the inner and outer rings:
\begin{equation}
    V_{\rm q}(r) = \Delta U \Theta(R_b - r),
\end{equation}
where $\Delta U = 0.01 U_{\rm b}$ is the quench amplitude. This weak perturbation is removed at $t=0$, after which the system evolves freely according to the Gross-Pitaevskii equation. The subsequent dynamics of the population imbalance $Z(t)$ is analyzed both in the time domain and through Fourier analysis to extract oscillation frequencies.
Fig.~\ref{fig:2_N_and_FFT} shows the results of this protocol for three values of applied acceleration: $a=0$, $a=10 \text{ mm/s}^2$, and $a=20 \text{ mm/s}^2$. In the symmetric case with $a=0$ (left column), the population imbalance exhibits a sinusoidal oscillation with a well-defined frequency, often termed the Josephson plasma frequency, which is characteristic of the Josephson effect in simple geometries. The corresponding Fourier spectrum displays a single sharp peak centered at this frequency.

As acceleration is applied $(a > 0)$, the time-domain oscillations reveal a qualitative change in behavior. The oscillation pattern becomes modulated, with a beating structure visible in the population imbalance. This beating reflects the presence of multiple frequency components in the Josephson spectrum, as clearly seen in the Fourier transforms (middle and right columns of Fig.~\ref{fig:2_N_and_FFT}). \textcolor{black}{The emergence of several distinct spectral peaks shows that the population-imbalance response is redistributed among multiple collective modes, with the multimode character becoming increasingly pronounced for growing acceleration.}

\subsection{Bogoliubov spectrum under symmetry breaking}

\textcolor{black}{The physical origin of the observed multimode response is intimately connected to the breaking of axial symmetry by the applied linear acceleration.}
To understand the microscopic mechanism underlying this multi-frequency Josephson dynamics, we now analyze the \textcolor{black}{Bogoliubov-de Gennes} spectrum of elementary excitations. The BdG equations [Eq.~(\ref{eq:BdG_equations})] are solved numerically for the same range of acceleration values, with the ground state $\psi_0$ obtained from a numerical solution of the GPE with no quench potential.

\begin{figure}
    \centering
    \includegraphics[width=\linewidth]{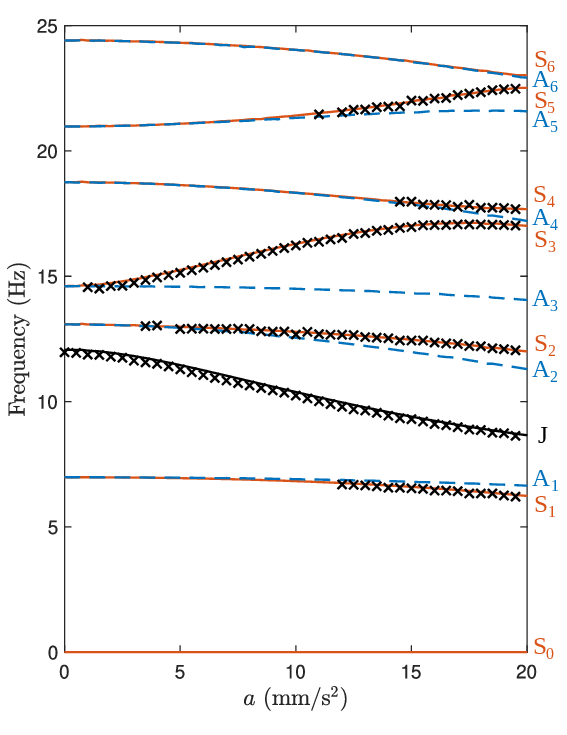}
    \caption{Elementary-excitation spectrum as a function of the applied acceleration $a$. Solid red lines represent modes symmetric with respect to the $x$ axis (labeled $S_n$), while dashed blue lines represent antisymmetric modes (labeled $A_n$). The zero-frequency Goldstone mode is labeled $S_0$. The mode labeled $J$ is the principal Josephson mode discussed in the text. For $a>0$, symmetry breaking activates additional modes in the population-imbalance dynamics, leading to the multiplet structure seen in Fig.~\ref{fig:2_N_and_FFT}. Black crosses denote the peak frequencies extracted from the dynamical simulations.}   \label{fig:3_BdG_JE_spectra_comparison}
\end{figure}

Figure~\ref{fig:3_BdG_JE_spectra_comparison} presents the dependence of the lowest Bogoliubov modes frequencies on applied acceleration. 
By comparing the mode frequencies with the peaks observed in the dynamics simulations (black crosses in Fig.~\ref{fig:3_BdG_JE_spectra_comparison}), we can see that only a subset of these modes can be identified within the observed spectrum. We also observe no apparent correlation between the frequencies of the modes and the minimal acceleration at which they become visible in the Josephson spectrum. \textcolor{black}{This indicates that the multimode response results from a spatial redistribution of the mode wave functions and from their modified coupling to the tunneling dynamics, rather than from simple shifts or mixing of modes with nearby frequencies.}

\begin{figure}
    \centering
    \includegraphics[width=\linewidth]{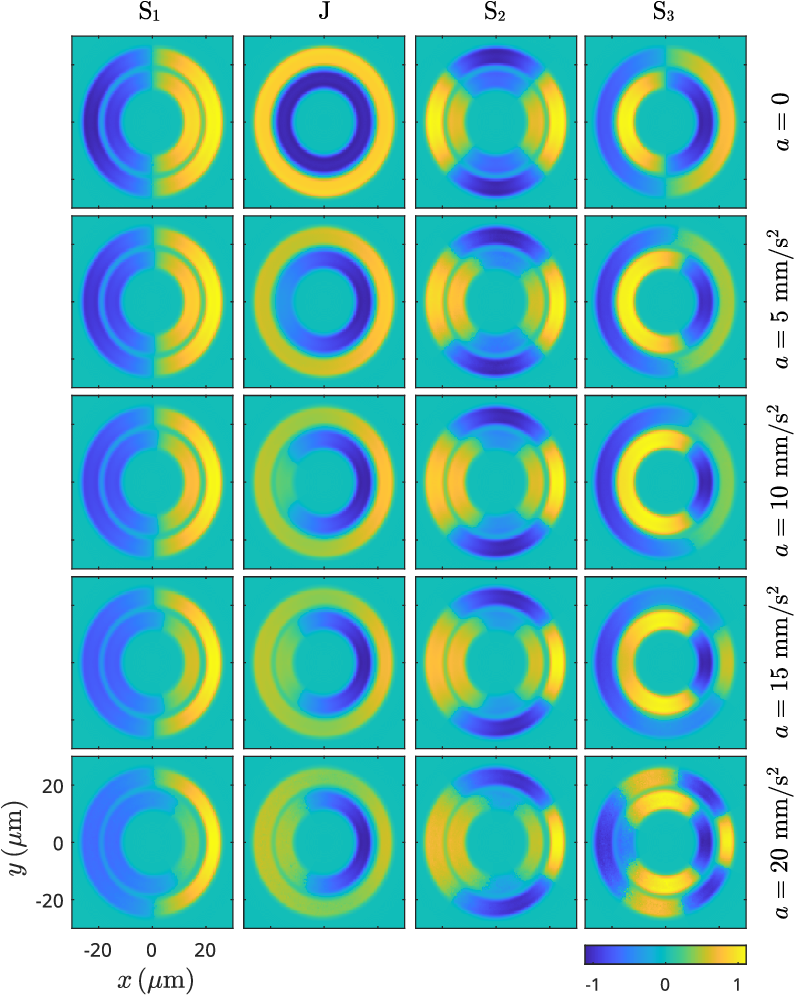}
     \caption{Spatial structure of the lowest Bogoliubov modes active in the population-imbalance dynamics, represented by the density perturbation $\delta\rho_n$ defined in Eq.~(\ref{eq:I_mode_density}). The rows correspond to $a=0$ (top), $5\,\mathrm{mm/s^2}$, $10\,\mathrm{mm/s^2}$, $15\,\mathrm{mm/s^2}$, and $20\,\mathrm{mm/s^2}$ (bottom). With increasing acceleration, the mode profiles become progressively distorted and asymmetric, consistent with the growth of their activity in Fig.~\ref{fig:mode_activity} and with the associated changes in the tunneling dynamics.}
    
    \label{fig:mode_spatial_structure}
\end{figure}

To get a better understanding of the relationship between the Bogoliubov spectrum and the Josephson dynamics, we analyze the spatial structure of the collective modes which are active in the Josephson oscillations. We define the density perturbation induced by a Bogoliubov mode $n$ as follows:
\begin{equation}
    \rho_n = |\psi_0|^2 + \delta\rho_n, \quad \delta\rho_n=2\text{Re}(\psi_0\, u_n^* + \psi_0 v_n).
    \label{eq:I_mode_density}
\end{equation}
The density perturbations associated with the lowest active modes are shown in Fig.~\ref{fig:mode_spatial_structure} for different values of acceleration.
In the absence of acceleration ($a=0$), the system possesses full rotational symmetry about the $z$ axis. All Bogoliubov modes can be labeled by a radial quantum number (distinguishing in-phase and out-of-phase oscillations between the rings) and an angular momentum quantum number. In this case, it is obvious that in-phase modes can not contribute to the Josephson dynamics, as they do not induce any current flow between the rings (see modes $S_1$ and $S_2$ in Fig.~\ref{fig:mode_spatial_structure}). Out-of-phase modes with non-zero angular momentum can in principle contribute to the Josephson dynamics, but their net contribution is suppressed due to the symmetry of their spatial structure, which leads to cancellation of tunneling flows in different angular sectors (see mode $S_3$ in Fig.~\ref{fig:mode_spatial_structure}). 

When acceleration is applied, the axial rotational symmetry is broken, leaving only reflection symmetry with respect to the plane containing the acceleration direction and the $z$ axis. Under this reduced symmetry, all modes must transform either as symmetric or antisymmetric with respect to this reflection. 
The degeneracies present at $a=0$ are lifted, and previously degenerate mode pairs split into distinct symmetric and antisymmetric components with different frequencies. Worth noting, the lowest Josephson mode (marked as $J$) is the only mode that does not split into two as acceleration increases.

This fundamental change in the symmetry structure has profound consequences for the role of modes in the Josephson dynamics. First, the above arguments based on the in-phase and out-of-phase character of modes no longer apply, as the modes can now have a mixed character and their spatial structure is distorted by the acceleration-induced potential. Also, the symmetry-based cancellation of tunneling flows is lifted for modes that are symmetric with respect to the $x$ axis, as angular sectors that contribute to tunneling flows in opposite directions have different sizes and therefore do not cancel each other. Finally, the number of nodes within the inner and outer rings is not necessarily the same for a given mode, which further modifies the structure of tunneling flows (see Fig.~\ref{fig:mode_spatial_structure}). 

As is seen in Fig.~\ref{fig:3_BdG_JE_spectra_comparison}, if we classify the modes according to their symmetry with respect to the $x$ axis as either $S$-type (symmetric with respect to the $x$ axis, \textcolor{black}{shown as solid red lines}) or $A$-type (antisymmetric, shown as dashed blue lines), it is clear that while all $A$-type modes remain essentially inactive across the entire acceleration range studied, \textcolor{black}{the low-lying $S$-type modes considered here generally become active in the Josephson dynamics as the acceleration increases.} 

\subsection{Mode symmetry and activity in Josephson oscillations}
Not all Bogoliubov modes contribute equally to the Josephson dynamics. A key observation from the comparison between the full excitation spectrum and the observed oscillations is that only a small subset of modes with specific properties significantly affects the population imbalance oscillations between the rings. \textcolor{black}{Thus, the relevant question is not only where the BdG eigenfrequencies lie, but whether the associated eigenvectors have a nonzero projection onto the experimentally measured imbalance channel.}

To characterize this selective coupling, we introduce the mode activity as a measure of a given mode's ability to induce population imbalance:
\begin{equation}
    D_n = \langle\psi_0|\,\hat{Z}\,|u_n+v_n\rangle,
    \label{eq:mode_activity}
\end{equation}
where $\hat{Z} = \Theta(r-R_b) - \Theta(R_b - r)$ is the population imbalance operator defined in the previous section and $u_n$, $v_n$ are the Bogoliubov mode amplitudes.

The physical meaning of mode activity is straightforward: a mode with large $|D_n|$ efficiently couples to tunneling flows between the rings, directly modulating the population imbalance. Conversely, modes with small activity $|D_n|$ produce density perturbations that are primarily confined to individual rings and do not significantly contribute to inter-ring tunneling.

\begin{figure}
    \centering
    \includegraphics[width=\linewidth]{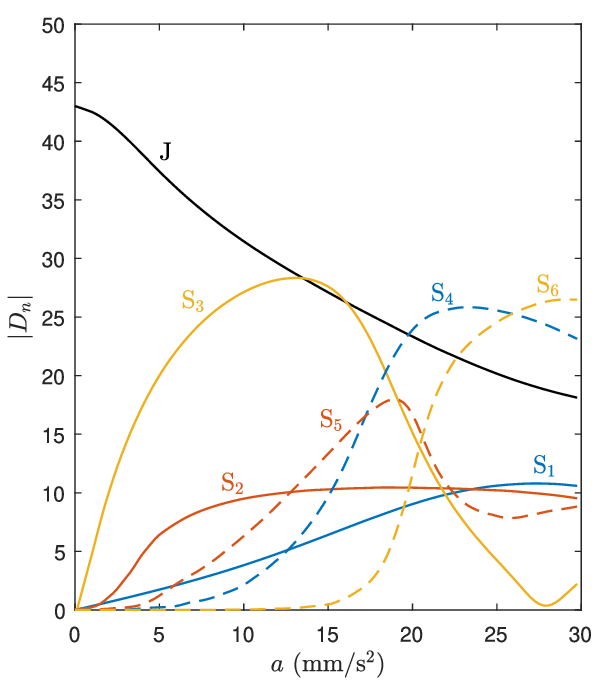}
    \caption{Activity $D_n$ of the different BdG modes as a function of the applied acceleration $a$. The Josephson mode (black line) maintains high activity across all accelerations. $S$-type modes (blue, red and yellow lines) increase their activity with increasing acceleration, while $A$-type modes (not shown) remain essentially inactive throughout. The activation of $S$-type modes directly corresponds to the appearance of additional components in the Josephson oscillation spectrum.}
    \label{fig:mode_activity}
\end{figure}

Figure~\ref{fig:mode_activity} shows the acceleration dependence of activity for the lowest lying modes. Several important features emerge from this analysis. First, the Josephson mode itself \textcolor{black}{(marked in black)} maintains relatively high activity across all acceleration values, consistent with its role as the primary driver of population imbalance oscillations. 
Second, nevertheless, the activity of this mode monotonically decreases with increasing acceleration. This can be qualitatively understood as a consequence of the distortion of the mode's spatial structure by the acceleration-induced potential, which reduces its overlap with the axially-symmetric population imbalance operator.
Third, higher modes have zero activity in the symmetric case ($a=0$) but gradually become activated as acceleration increases. This activation of higher modes is closely correlated with the appearance of additional components in the Josephson frequency spectrum (Fig.~\ref{fig:2_N_and_FFT}) at finite acceleration values.

Importantly, all analysed $S$-type modes systematically show growing activity as acceleration increases, while $A$-type modes remain essentially inactive across the entire acceleration range studied. This symmetry-based selection rule reflects the underlying structure of the ground state and its response to the acceleration perturbation.

\textcolor{black}{We note that the predominantly single-frequency imbalance dynamics observed in the higher-barrier regime of Ref. \cite{PhysRevA.111.043308} is consistent with the present interpretation: higher BdG modes are expected to have only weak activity when their density perturbations are largely localized within individual rings, resulting in a small overlap with the population-imbalance channel.}

\textcolor{black}{The results presented in this subsection establish that the observed multimode Josephson response arises from the interplay between symmetry breaking, mode activation, and mode-weighted coupling to tunneling dynamics. It is not a simple upward or downward shift of a single spectral peak, but rather a redistribution of oscillator strength among multiple collective modes, with new modes becoming activated as the acceleration increases.}
This provides a physical foundation for using Josephson spectroscopy as a sensitive probe of symmetry-breaking perturbations and collective excitations in ring-shaped condensate systems.

\section{Resonant driving of Josephson oscillations and probing of Bogoliubov modes}

In the previous section we have shown how external symmetry breaking potentials can influence the spatial structure of the Bogoliubov modes and what impact this has on the measured Josephson oscillations. We now go one step further and consider how the spatial structure of these modes can be probed experimentally by means of an additional localized time-dependent perturbation.

We consider now the BEC initially in the stationary state $\psi_0$ exposed at $t > 0$ to a small periodic driving potential $\delta V(\mathbf{r}) \sin \omega_d t$, where the spatial profile of the perturbation is given by: 
\begin{equation}\label{eq:perturbation}
   \delta V(\mathbf{r}) = \delta U \,\exp\left[-\frac{1}{2 l_d^2} (x\sin{\varphi_d} - y\cos{\varphi_d})^2\right]
\end{equation}
\textcolor{black}{This perturbation is localized in the angular coordinate around the direction defined by the angle $\varphi_d$ (measured from the \textcolor{black}{$x$ axis}) and has the width $l_d$ in the azimuthal direction (see Fig.~\ref{fig:1_setup}(a)). 
This choice reflects the experimentally most accessible implementations of stirring potentials in ring systems \cite{PhysRevLett.110.025302, PhysRevA.91.033607}.}
The amplitude of this potential is assumed to be small $\delta U \ll \mu$ so that linearized treatment is applicable. Unless otherwise stated, the results presented below are obtained for the following parameters of the perturbation: $\delta U = 0.01 U_\mathrm{b}$, $l_d = 4\,\mu$m, $\varphi_d = 0$.

\begin{figure}[tbp]
    \centering
    \includegraphics[width=\linewidth]{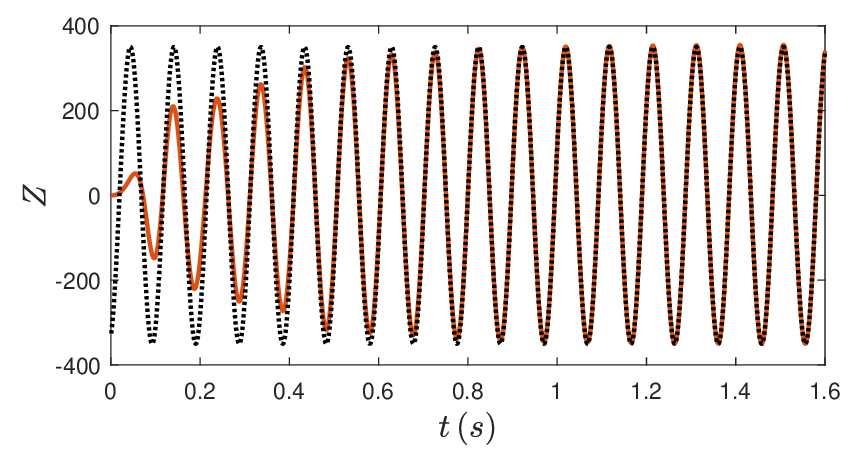}
    \caption{
        Oscillations of the population imbalance $Z(t)$ between the rings under periodic driving with frequency $\omega_{d}/2\pi = 10.25\,$Hz close to the resonance with one of the Bogoliubov modes. 
        The dissipation parameter $\gamma=5\times10^{-3}$ stabilizes the oscillation amplitude after $\sim 0.8\,$s.
        Solid red line shows the result of the numerical solution of the dissipative GPE (\ref{eq:diss_GPE_perturbation}), while the dotted black line is a sinusoidal fit $Z(t) = Z_0 \sin(\omega_d t + \delta)$ performed for $t>0.8\,$s.
    }\label{fig:8_dissipative_driving}
\end{figure}

\textcolor{black}{To perform a quantitative analysis of the system response to the
periodic perturbation, we start from the dissipative Gross-Pitaevskii
equation including the driving term,
\begin{equation}\label{eq:diss_GPE_perturbation}
    (i-\gamma)\hbar\frac{\partial\psi}{\partial t}
    =
    \big(
    \hat{\mathcal{H}}_0 - \mu
    + g|\psi|^2
    + \delta V \sin{\omega_{d} t}
    \big)\psi .
\end{equation}
Here $\gamma$ is a phenomenological dissipation parameter
\cite{PhysRevA.57.4057}. It is introduced to describe weak
energy relaxation at the mean-field level and to regularize the
resonant response: in the absence of damping, a drive exactly
resonant with a Bogoliubov mode would lead, within linear theory,
to the secular growth of the oscillation amplitude. For finite
$\gamma$, the initial transients decay and the system approaches
a steady oscillatory regime with a finite amplitude and a well-defined
phase lag, as illustrated in Fig.~\ref{fig:8_dissipative_driving}.
Our goal is to analyze this frequency response and, in particular,
to determine how it can be read out from the population-imbalance
oscillations between the rings.}

The wave function $\psi$ is represented as a sum of the stationary state $\psi_0$ and a small perturbation $\delta\psi$:
\begin{equation}
    \psi(\mathbf{r},t) =  \psi_0(\mathbf{r}) + \delta\psi(\mathbf{r},t)
\end{equation}
The perturbation $\delta\psi$ then satisfies the following linearized equation:
\begin{equation}\label{eq:delta_psi_equation}
(i-\gamma) \hbar\, \partial_t \delta\psi = \hat{\mathcal{L}} \delta\psi + g \psi_0^2 \delta\psi^* + \psi_0 \delta V \sin{\omega_{d} t},
\end{equation}
where $\hat{\mathcal{L}}$ is defined in Eq.~(\ref{eq:L_operator}).
The perturbation $\delta\psi$ can be written as a superposition of the Bogoliubov modes of the stationary state $\psi_0$ \cite{PhysRevA.72.023613,PhysRevA.99.023619}:
\begin{equation}\label{eq:delta_psi_expansion}
    \delta\psi(\mathbf{r},t) = \sum_n \big( \alpha_n (t) u_n(\mathbf{r}) + \alpha^*_n (t) v_n^*(\mathbf{r}) \big).
\end{equation}
To simplify the analysis, in what follows we focus on the stabilized regime of oscillations, when the initial transient processes have already decayed due to dissipation and the system oscillates at the driving frequency $\omega_{d}$ (see Fig.~\ref{fig:8_dissipative_driving}).
Linearized treatment also implies that we can restrict ourselves to the first Floquet band of the time-periodic equation (\ref{eq:delta_psi_equation}) and write the time-dependent coefficients $\alpha_n(t)$ in the form:
\begin{equation}\label{eq:alpha_n_expansion}
    \alpha_n = A_n e^{-i\omega_{d} t} +  B_n e^{i\omega_{d} t}.
\end{equation}
Substituting the expansion (\ref{eq:delta_psi_expansion}, \ref{eq:alpha_n_expansion}) into the dissipative GPE (\ref{eq:diss_GPE_perturbation}) and projecting onto the eigenfunctions of the BdG equations, we obtain a linear system of equations for the coefficients \textcolor{black}{$A_n$ and $B_n$:
\begin{equation}\label{eq:An_Bn_equations}
    \left\{
    \begin{aligned}
         \hbar (\omega_{d}-\omega_n) A_n  + \sum\limits_m i \gamma \hbar\omega_{d} \beta_{nm} A_m \\
         + \sum\limits_m i \gamma \hbar \omega_{d} \varepsilon_{nm} B_m = F_n/2i, \\[5pt]
        \hbar (\omega_{d}+\omega_n) B_n  - \sum\limits_m i \gamma \hbar\omega_{d} \beta_{nm} B_m \\
         - \sum\limits_m i \gamma \hbar \omega_{d} \varepsilon_{nm} A_m = F_n/2i,
    \end{aligned}
    \right.
\end{equation}}
where
\begin{gather}
    \beta_{nm} = \int d\mathbf{r} \Big( u_n^* u_m + v_n^* v_m \Big), \\
    \varepsilon_{nm} = \int d\mathbf{r} \Big( u_n^* v_m^* + v_n^* u_m^* \Big), \\
    F_n = \langle \psi_0 | \delta V | u_n^* + v_n^* \rangle = \int d\mathbf{r} \, \psi_0 \delta V \Big( u_n^* + v_n^* \Big),
\end{gather}
assuming that the stationary state $\psi_0$ is real-valued.

Further simplification of the system (\ref{eq:An_Bn_equations}) can be achieved by noting that the dissipation term is small $\gamma \ll 1$ as well as the off-diagonal terms in $\beta_{nm}$ and $\varepsilon_{nm}$, and therefore the coupling between different modes due to dissipation can be safely neglected. 
The remaining system of two coupled equations for the amplitudes $A_n$ and $B_n$ of the $n$-th mode can be easily solved, yielding the following expressions:
\begin{gather}
    A_n = \frac{F_n}{2i\hbar}\frac{\omega_{d}+\omega_n - i \gamma \omega_{d} (\beta_{nn}+\varepsilon_{nn})}{ \omega_{d}^2 - \omega_n^2 + 2 i \gamma \omega_{d} \omega_n \beta_{nn}}, \label{eq:amp_A}\\
    B_n = \frac{F_n}{2i\hbar}\frac{\omega_{d}-\omega_n + i \gamma \omega_{d} (\beta_{nn}+\varepsilon_{nn})}{ \omega_{d}^2 - \omega_n^2 + 2 i \gamma \omega_{d} \omega_n \beta_{nn}}, \label{eq:amp_B}
\end{gather}
where we additionally neglected the terms proportional to $\gamma^2$ in the denominators. These expressions clearly show the resonant behavior of the amplitude $A_n$ at the frequencies of the Bogoliubov modes $\omega_{d} \approx \omega_n$, where the denominators become small and the amplitude reaches its maximum value. \textcolor{black}{Since in the vicinity of the resonance $A_n \sim (\omega_{d}-\omega_n)^{-1}$, the amplitude $A_n$ is expected to be much larger than $B_n \sim (\omega_{d}+\omega_n)^{-1}$.} Therefore in practically relevant frequency ranges we may completely neglect $B_n$ in Eq.~(\ref{eq:An_Bn_equations}) and write the following approximate equations for the amplitudes $A_n$ of the Bogoliubov modes excited by the periodic perturbation:
\begin{equation}
    A_n = \frac{1}{2i\hbar}\frac{F_n}{\omega_{d}-\omega_n + i \gamma \omega_{d} \beta_{nn}}.
\end{equation}
The results below are obtained using the full expressions for $A_n$ and $B_n$ given by Eqs.~(\ref{eq:amp_A}) and (\ref{eq:amp_B}), but the simplified expression above can be used to obtain a qualitative understanding of the system response in the resonant regime. 

Having determined the amplitudes $A_n$ and $B_n$ of the excited Bogoliubov modes, we can calculate the population imbalance between the rings as follows (see Eq.~(\ref{eq:population_imbalance_operator})):
\begin{multline}
    Z(t) = \langle \psi | \hat Z | \psi \rangle  - \langle \psi_0 | \hat Z | \psi_0 \rangle 
    = \langle \psi_0 | \hat Z | \delta\psi \rangle + \langle \delta\psi | \hat Z | \psi_0 \rangle \\ = 2 \text{Re}  \sum_n \left[ \left(A_n e^{-i\omega_{d} t} + B_n e^{i\omega_{d} t}\right) \langle \psi_0 | \hat Z | u_n \rangle \right. \\ \left. + \left( A_n^* e^{i\omega_{d} t} + B_n^* e^{-i\omega_{d} t} \right) \langle \psi_0 | \hat Z | v_n^* \rangle \right].
\end{multline}
From this expression we can obtain the amplitude and phase of the population imbalance oscillations $Z(t) = Z_0 \sin (\omega_{d} t + \delta)$, which evaluate to the following expressions:
\begin{gather}
Z_0 = 2 \Big |\sum_n  \big( A_n D_n + B_n^* D_n^* \big) \Big| \label{eq:z0_analytical}\\
\delta = \frac{\pi}{2}-\text{arg} \left( \sum_n \big( A_n D_n + B_n^* D_n^* \big) \right), \label{eq:delta_analytical}
\end{gather}
where $D_n$ is the activity of the $n$-th mode defined in Eq.~(\ref{eq:mode_activity}). The values obtained from Eqs.~(\ref{eq:z0_analytical}) and (\ref{eq:delta_analytical}) can be directly compared with the numerical simulations of the GPE and used to extract information about the frequencies and spatial structure of the Bogoliubov modes of the system.

\begin{figure*}[htb]
    \centering
\includegraphics[width=0.9\linewidth]{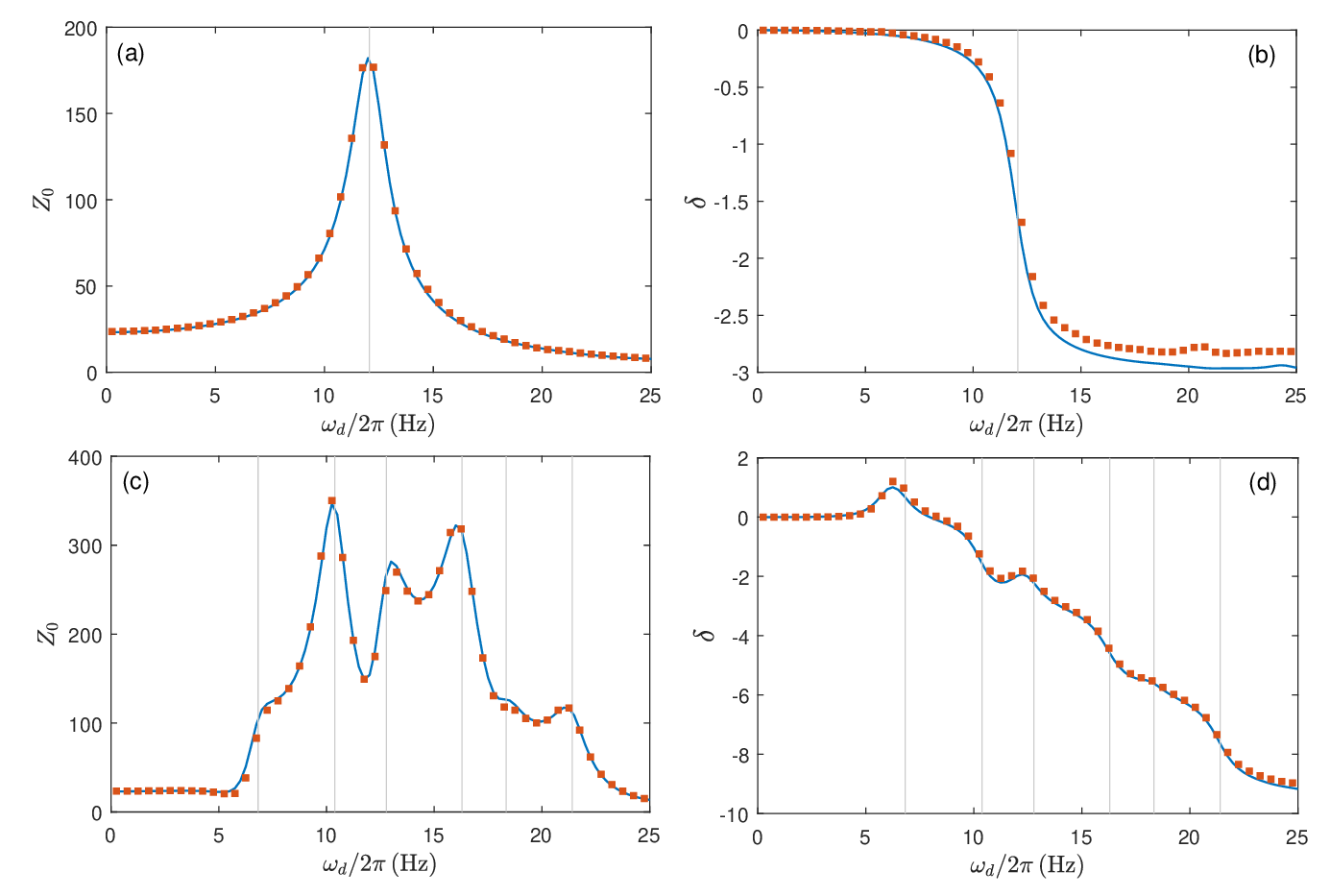}
    \caption{
        Amplitude $Z_0$ [panels (a) and (c)] and phase $\delta$ [panels (b) and (d)] of the population imbalance oscillations as a function of the excitation frequency $\omega_{d}$ for the axisymmetric case with $a=0$ [panels (a) and (b)] and the non-axisymmetric case with $a=10$ mm/s$^2$ [panels (c) and (d)]. Red squares show the results obtained from the numerical simulations of the GPE, while the solid blue lines are the theoretical predictions from Eqs.~(\ref{eq:z0_analytical}) and (\ref{eq:delta_analytical}). The vertical gray lines indicate the frequencies of the active Bogoliubov modes at the corresponding acceleration.
}\label{fig:9_amplitude_phase_comparison}
\end{figure*}

In Fig.~\ref{fig:9_amplitude_phase_comparison} we present the results of such comparison for the axisymmetric case with $a=0$ and the non-axisymmetric case with $a=10$ mm/s$^2$. The amplitude and phase of the population imbalance oscillations obtained from the numerical simulations of the GPE are shown as a function of the excitation frequency $\omega_{d}$ and compared with the theoretical predictions derived above. The amplitude and phase were extracted from the numerical data by fitting the population imbalance oscillations in the stabilized regime with a sinusoidal function $Z(t) = Z_0 \sin (\omega_{d} t + \delta)$, as shown in Fig.~\ref{fig:8_dissipative_driving}. 

We can see from Fig~\ref{fig:9_amplitude_phase_comparison} that in both symmetric and symmetry-broken cases, the theoretical model captures the resonant behavior of the system well, with peaks in the amplitude corresponding to the frequencies of the Bogoliubov modes, and the phase showing characteristic shifts around these resonances. The correspondence between the analytical predictions and numerical results is very good, justifying the assumptions made in the derivation and confirming the validity of the theoretical model in describing the response of the system to periodic perturbations. It is worth noting that the frequency response curves do not show any peaks at the frequencies of the modes that are inactive in the Josephson spectrum (see Fig.~\ref{fig:3_BdG_JE_spectra_comparison}), which means that the proposed protocol can be used to selectively probe only the modes contributing to the Josephson dynamics.

\begin{figure}[tb]
    \centering    \includegraphics[width=\linewidth]{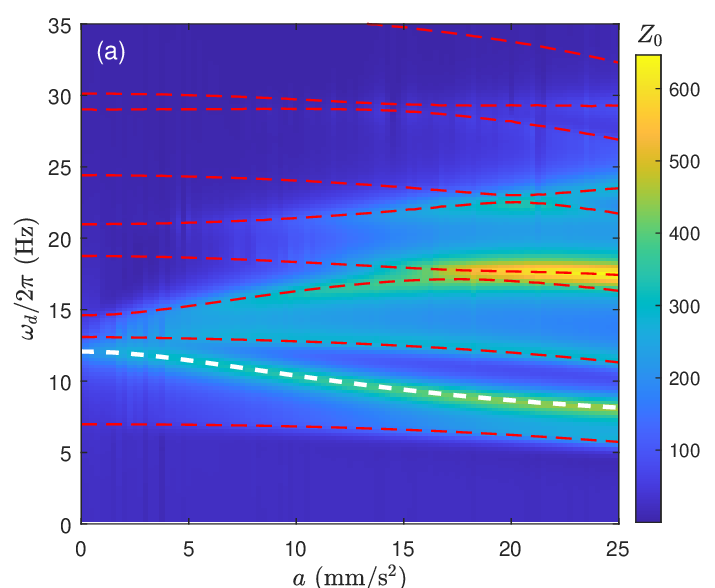}
\includegraphics[width=\linewidth]{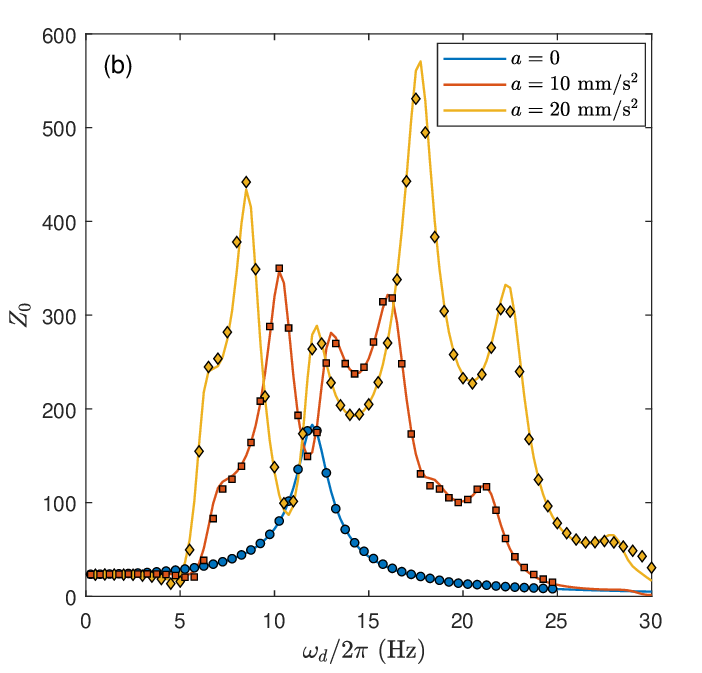}
    \caption{Population imbalance as a function of the applied acceleration $a$ and the drive frequency $\omega_{d}$. (a) The color map shows the amplitude $Z_0$ of the population imbalance oscillations as a function of the excitation frequency $\omega_{d}$ and the applied acceleration $a$. Red dashed lines indicate the frequencies of $S$-type Bogoliubov modes, while the white dashed line is the $J$ mode. (b) The panel shows the frequency response curves for three different values of acceleration: $a=0$ (blue line), $a=10$ mm/s$^2$ (red line) and $a=20$ mm/s$^2$ (yellow line). Symbols of corresponding colors are the results obtained from numerical simulations of the GPE.} 
    \label{fig:10_waterfall}
\end{figure}

In Fig.~\ref{fig:10_waterfall} we present in more detail the dependence of the frequency response curve on the applied acceleration. We can see that as the acceleration increases, frequencies of the Bogoliubov modes shift and additionally, more modes become active in the Josephson spectrum, which leads to the appearance of additional components in the frequency response curve. The amplitude of resonance peaks also increases with the acceleration, which interestingly does not explicitly follow the increase or decrease of the mode activity $D_n$ with acceleration, but is also influenced by the spatial structure of the modes and their overlap with the perturbation $\delta V$ (see Eq.~(\ref{eq:An_Bn_equations})). In other words, since the perturbation is localized at $\varphi_d=0$, the observed increase of the oscillation amplitude indicates the increase of the mode density at this angle, which correlates with the decrease of the ground state density at this angle due to the acceleration action.

\textcolor{black}{From Eqs.~(\ref{eq:amp_A}), (\ref{eq:amp_B}), and
(\ref{eq:z0_analytical}) one can see that the spatial profile
of the driving perturbation $\delta V$ enters the response only
through the overlap integrals $F_n$. Near an isolated resonance,
where a single Bogoliubov mode gives the dominant contribution,
the amplitude of the population-imbalance oscillations is
therefore controlled by $|F_n|$. This observation extends the
frequency-domain Josephson spectroscopy discussed above: after
selecting an active mode by tuning the driving frequency to its
resonance, one can scan the position of the localized perturbation
and use the resulting response amplitude to probe the mode
profile.}

\begin{figure}[tb]
    \centering
    \includegraphics[width=\linewidth]{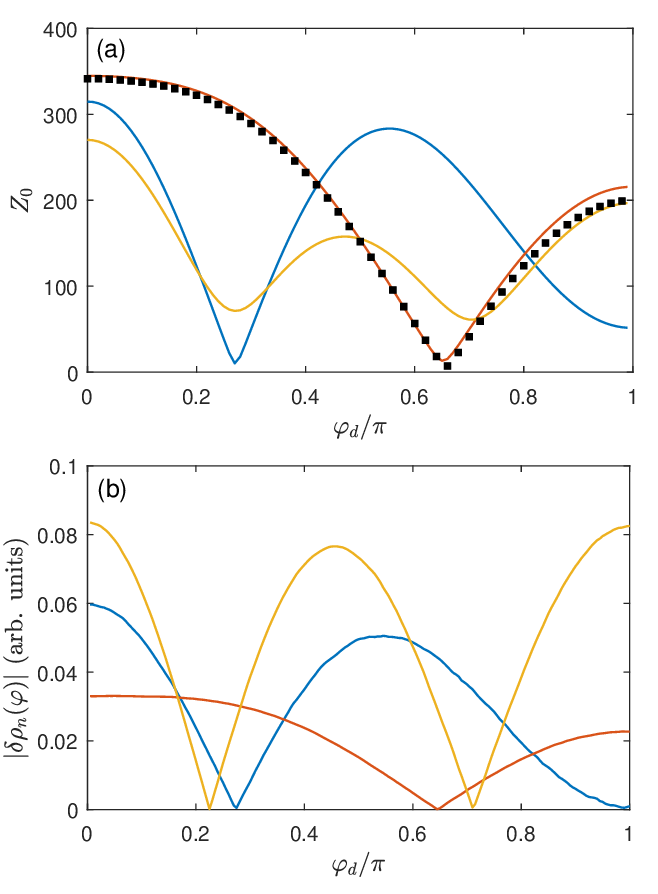}
    \caption{
        Panel (a) shows the angular dependence of the amplitude of population imbalance oscillations $Z_0$ for three different frequencies of the periodic perturbation $\omega_{d}$ corresponding to the resonances with different Bogoliubov modes: $\omega_{d}/2\pi = 10.3\,$Hz (red line), $\omega_{d}/2\pi = 12.8\,$Hz (yellow line) and $\omega_{d}/2\pi = 16.3\,$Hz (blue line). \textcolor{black}{Acceleration magnitude is $a=10$ mm/s$^2$.} Black squares show for comparison the result obtained from numerical simulation of the GPE with the applied driving frequency $\omega_{d}/2\pi = 10.3\,$Hz, which is in good agreement with the prediction for the respective Bogoliubov mode. Panel (b) shows the angular dependence of the Bogoliubov modes corresponding to these resonances, quantified by the density perturbation $\delta\rho_n = 2\text{Re}(\psi_0 u_n^* + \psi_0 v_n)$, integrated over the radial coordinate and plotted as a function of the angular coordinate $\varphi$. The figure shows the absolute value for a better comparison with panel (a).
    }
\label{fig:resonance_angular_dependence}
\end{figure}

\textcolor{black}{
To illustrate this idea,
Fig.~\ref{fig:resonance_angular_dependence}(a) shows the
resonance amplitude of the population imbalance as a function
of the angular position $\varphi_d$ of the localized periodic
perturbation. This angular dependence is compared with the
structure of the corresponding Bogoliubov modes, shown in
Fig.~\ref{fig:resonance_angular_dependence}(b) through the
radially integrated density perturbation
$\delta\rho_n=2\mathrm{Re}(\psi_0 u_n^*+\psi_0 v_n)$ plotted as a
function of the angular coordinate $\varphi$. The correspondence
is not expected to be exact in all cases, because nearby modes
may still contribute appreciably even at resonance, as seen for
the intermediate mode shown by the yellow line. However, for the
other two modes the agreement is very good, demonstrating that
the angular scan provides a spectroscopic fingerprint of the
selected mode. These results show that the proposed
Josephson-spectroscopy protocol provides both frequency-domain
identification of active Bogoliubov modes and access to their
angular structure.}

\section{Conclusions}
\textcolor{black}{
We have investigated Josephson dynamics in a long atomic
Bose-Josephson junction formed by two tunnel-coupled
coplanar $^{87}$Rb condensate rings under an in-plane linear
acceleration. Combining GPE simulations with
BdG analysis, we have shown that
acceleration-induced symmetry breaking qualitatively changes
the population-imbalance dynamics. In the axially symmetric
case the response is dominated by a single Josephson plasma
oscillation, whereas for finite acceleration it develops a
multi-component spectrum that is visible both as beating in the time-domain dynamics and as a multiplet in the Fourier spectrum.}

\textcolor{black}{
This spectral multiplet should not be interpreted simply as a
splitting of one Josephson mode. Rather, it results from the
activation of Bogoliubov modes that acquire finite overlap with
the population-imbalance operator when rotational symmetry is
reduced to reflection symmetry. 
The spectral peaks observed in the GPE simulations coincide with
the corresponding active BdG eigenfrequencies, while the calculated
mode activities explain why only selected modes contribute to the
Josephson response. Modes symmetric with respect to the acceleration
direction acquire finite
activity in the imbalance dynamics, whereas antisymmetric modes
remain essentially inactive and do not couple to
population-imbalance oscillations. }

\textcolor{black}{
We have also proposed a Josephson-spectroscopy protocol based
on a weak localized periodic perturbation. By scanning the
driving frequency, active Bogoliubov modes are detected as
resonances in the amplitude and phase of the population-imbalance
oscillations. A dissipative time-dependent BdG theory provides
analytical response functions in terms of the mode activity
$D_n$, the drive--mode overlap $F_n$, and the damping parameter
$\gamma$. These functions quantitatively agree with full
GPE simulations in the linear-response regime and
clarify how the resonance lineshapes are controlled by the mode
activity, the drive-mode overlap, and damping. }

\textcolor{black}{
Finally, we have shown that the resonant response can be used
to probe the spatial structure of active modes. When the drive
frequency is tuned near a spectrally isolated resonance, scanning
the angular position of the localized perturbation makes the
response amplitude reflect the angular structure of the
corresponding mode density perturbation, with the resolution set
by the drive profile and by possible contributions from nearby
modes. This angular response can therefore serve as a fingerprint
for identifying active Bogoliubov modes.}

\textcolor{black}{
In summary, the free-evolution and resonantly driven protocols
provide complementary information about the collective excitation
spectrum: the former reveals which modes enter the natural
Josephson dynamics, while the latter allows their frequency and
spatial response to be probed in a controlled way. The strong
dependence of the active-mode spectrum on acceleration also
suggests a possible route toward symmetry-sensitive inertial
probing, while similar ideas could be extended to other controlled
symmetry-breaking perturbations, such as tilted potentials or
rotating frames. Our results identify acceleration-biased
dual-ring condensates as a controllable platform for
symmetry-selected Josephson dynamics and spectroscopic probing
of collective excitations in atomtronic circuits.}

\begin{acknowledgments} 
O.P. acknowledges support from the National Research Foundation of Ukraine grant (2023.03/0097) ``Electronic and transport properties of Dirac materials and Josephson junctions''. M.E.\,is supported by U.S.\,National Science Foundation Grant No.\,PHY-2207476.
\end{acknowledgments}

\bibliography{ref}  
\end{document}